\documentclass[conference,10pt,letterpaper]{IEEEtran}
\IEEEoverridecommandlockouts
\usepackage{amsmath,amssymb,amsfonts}
\usepackage{acronym}
\usepackage{algorithmic}
\usepackage{graphicx}
\usepackage{multirow} % model_qualification_table.tex spans Device rows
\usepackage{textcomp}
\usepackage{xcolor}
\usepackage[hidelinks]{hyperref}
\usepackage{acronym}
\usepackage{cleveref}
\crefname{figure}{Fig.}{Figs.}
\Crefname{figure}{Fig.}{Figs.}

\usepackage[backend=biber,,citestyle=numeric-comp,bibstyle=ieee,sorting=none,minbibnames=1,maxbibnames=1,doi=false,url=false,eprint=false]{biblatex}
\renewcommand{\thesection}{\textbf{\Roman{section}}}
\AtEveryBibitem{%
  \clearfield{title}%
  \clearfield{subtitle}%
}

\begin{document}

\title{\fontsize{24}{24}\selectfont{12 nm FinFETs from 292 K to 10 mK: Characterization and a Temperature-Continuous Compact Model for Qubit Control}}
\author{\fontsize{11}{11}\selectfont D. Lee\textsuperscript{1*}, J. Lee\textsuperscript{1*}, L. Jin\textsuperscript{1*}, R. Rangaraj\textsuperscript{1}, L. Shao\textsuperscript{1}, and J. S. Walling\textsuperscript{1}\\
\fontsize{10}{12}\selectfont \textsuperscript{1}Bradley Department of Electrical and Computer Engineering, Virginia Tech, Blacksburg, VA, USA, email: \href{mailto:dklee29@vt.edu}{dklee29@vt.edu}\\
\fontsize{10}{12}\selectfont \textsuperscript{*}Equally credited authors\vspace{-1ex}
}

\maketitle
\newacro{PDK}{process development kit}
\newacro{GF}{GlobalFoundries}
\newacro{RF}{radio frequency}
\newacro{DAC}{digital-to-analog converter}
\newacro{RF-DAC}{\ac{RF}-\ac{DAC}}
\newacro{SS}{sub-threshold slope}
\newacro{RVT}{regular-voltage-threshold}
\newacro{mK}{millikelvin}
\newacro{$V_{th}$}{threshold voltage}
\newacro{PCB}{printed circuit board}
\newacro{SMU}{source measurement unit}
\newacro{SoC}{system-on-chip}
\begin{abstract}
We report cryogenic characterization and compact modeling of \ac{GF} 12LP \ac{RVT} n- and p-FinFETs, measured from 292~K to 10~mK. Both devices retain sub-Kelvin gate control without carrier freeze-out, with the n-type device exhibiting an intrinsic \ac{SS} floor of 24.8~mV/dec. To bypass high on-chip series routing resistance that confounds peak-$g_m$ extraction, we adopt a fixed-current ($3.26~\mu\mathrm{A}/\mu\mathrm{m}$) methodology.  These low-current metrics inform a continuous 1.36~K to 292~K compact model using dynamic parameter injection, preserving room-temperature \ac{PDK} qualification.
\end{abstract}

\section{Introduction}

Scaling quantum processors toward fault-tolerant operation requires generating and routing large numbers of precisely timed, high-fidelity \ac{RF} control signals to qubit arrays held at \ac{mK} temperatures. Today, every qubit control channel requires a dedicated coaxial cable running from room-temperature electronics through successive cryostat stages, creating a severe wiring bottleneck that restricts scalability in both cable count and thermal budget~\cite{Enz2020_iedm, Qian2024_iedm}. Moving the \ac{RF-DAC} function directly inside the cryostat, at $<4\rm~K$, allows a single high-bandwidth digital bus to replace analog coaxial lines~\cite{Ram2024_iedm,Guo2024_tcasii,Fakkel2026_isscc}. Utilizing optical interconnects to drive a \ac{SoC} with an \ac{RF-DAC} can further minimize both heating and electromagnetic interference.  Realizing an \ac{RF-DAC} below 4~K places strict limits on power dissipation and understanding of device performance. Because cooling power at 10~\ac{mK} is limited to $\mu\mathrm{W}$, transistors, operating as RF switches presents an opportunity to minimize self-heating. \ac{GF} 12LP (12 nm FinFET) offers reduced parasitic capacitance and a favorable power-delay product for this application. Cryo-CMOS compact modeling has advanced~\cite{Incandela2018_jeds,Enz2020_iedm,Singh2022_edl,Gupta2023_ted,Sharma2025_jeds}, including modeling of \ac{GF} 12LP FinFETs to $<10\mathrm{K}$~\cite{Sarkar2024_laedc,12lpref}; however, those are narrow single-finger structures rather than the wide multi-finger geometries required for integration in \acp{RF-DAC}.  We offer an extension into the \ac{mK} regime through systematic experimental characterization of wide, multi-finger \ac{RVT} 12LP devices. In this work, we characterize \ac{GF} 12LP \ac{RVT} n- and p-FinFET devices from 292~K down to 10~\ac{mK}. The coldest points of the main acquisition are 0.404~K for the PFET and 0.633~K for the NFET, with a separate sub-20~mK acquisition extending the transfer characteristics to 10~\ac{mK}. Because long thin metal interconnect inflates the relative impact of series resistance in the source and drain routing (e.g., $R_{D,S}$), we evaluate the devices using low-current, weak-to-moderate inversion methodologies. Specifically, \ac{$V_{th}$} is extracted via a constant-current method at $1~\mu\mathrm{A}/\mu\mathrm{m}$. \ac{SS} is evaluated via the voltage difference across a full-decade of current in the subthreshold region. We demonstrate that traditional peak-$g_m$ extraction at these temperatures is limited by series resistance. Because cooling increases $g_m$ faster than it decreases resistance in the source and drain routing, $R_{D,S}$, the $g_m\times R_S$ product systematically rises, approaching the ceiling at which external $g_m$ saturates at $1/R_S$. Evaluating $g_m$ at a fixed $3.26~\mu\mathrm{A}/\mu\mathrm{m}$ holds the extraction clear of that ceiling and narrows the apparent N/P spread. Finally, we present a temperature-scalable compact model from 1.36~K to 292~K. By utilizing dynamic parameter injection and interconnect de-embedding, the model captures deep-cryo \ac{SS} saturation and a resistance floor without modifying the nominal room-temperature \ac{PDK}. 
\section{FinFET Measurement and Deembedding}
\begin{figure*}[htbp]
    \centering
    % 0.85 rather than full width: the artboard is 20.3 x 11.8 in, so at
    % \textwidth it is the tallest item in the block at 4.2 in. Trimming it
    % is what buys the figure block its second page back, and a setup
    % diagram tolerates scaling better than any of the data figures.
    \includegraphics[width=1\textwidth]{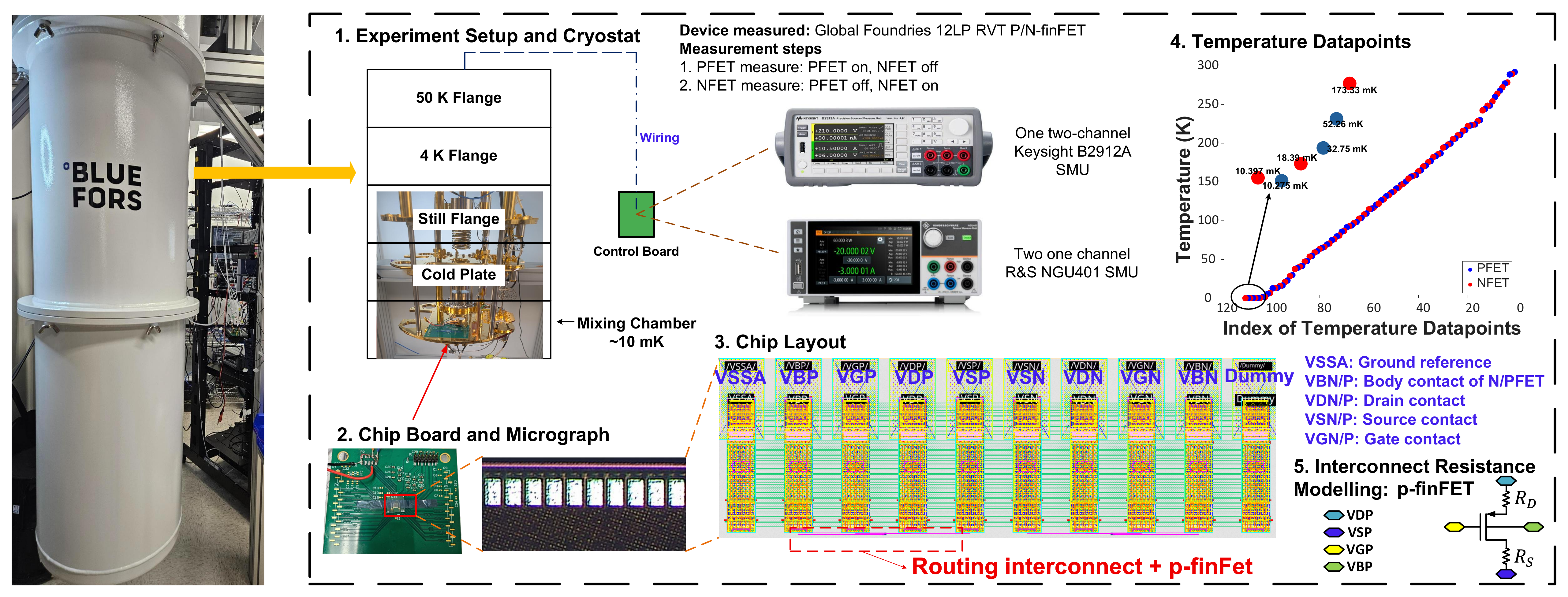}
    \caption{1. Experimental setup: The DUT sits in the 10mK mixing stage of a BlueFors dilution refrigerator. 2. The PCB and die photo. 3. The chip layout of the DUT. 4. The temperature data points by device. 5. Interconnect resistance $R_{S}$ and $R_{D}$ of the devices.}
    \label{fig:setup}
\end{figure*}

Both the \ac{RVT} n- and p-FinFETs were designed and laid out in the \ac{GF} 12LP process. Each device has four separate pads, one for each of the gate, source, drain, and body contact (\Cref{fig:setup}.3). The physical separation of the pads requires relatively long routing to interconnect the devices, resulting in relatively high $R_{D,S}$. The die was mounted to a custom \ac{PCB} with SMA connectors for connection to the measurement apparatus (\Cref{fig:setup}.2).

\noindent \textbf{Test Setup} The \ac{PCB} is located inside the BlueFors 10~\ac{mK} mixing chamber flange. Coaxial cables connect the terminals (S, G, D, B) of the devices from the inside to the outside of the chamber, which is then connected to a \ac{SMU} (\Cref{fig:setup}.1). The n- and p-FinFETs characteristics were measured alternately while the other was disabled. 

The devices were measured over the temperature range from 10~mK to 292~K during both a cooling and warming cycle. Those presented here are from the warming cycle and consist of $\approx50$ temperature points for each FET, sampled most densely between 50~K and ambient and thinning toward sub-Kelvin temperatures; the data points are shown in \Cref{fig:setup}.4. 

\begin{figure*}[htbp]
    \centering
    \includegraphics[width=\textwidth]{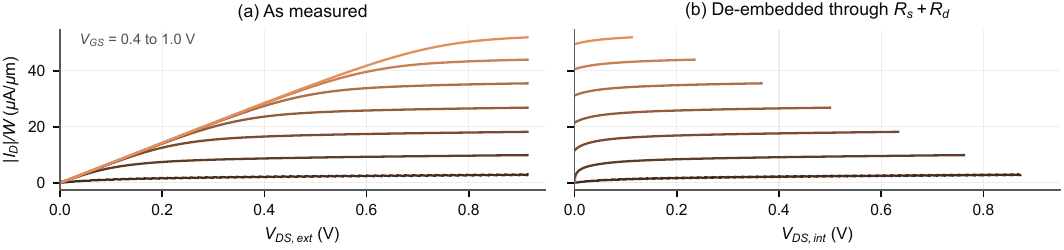}
    \caption{Output characteristics of the NFET at 5.10~K for
    $|V_{GS,ext}|$ from 0.4 to 1.0~V. (a)~Measured data before de-embedding routing resistance. (b)~Measured data after de-embedding routing resistance.}
    \label{fig:output}
\end{figure*}

\begin{figure*}[htbp]
    \centering
    \includegraphics[width=\textwidth]{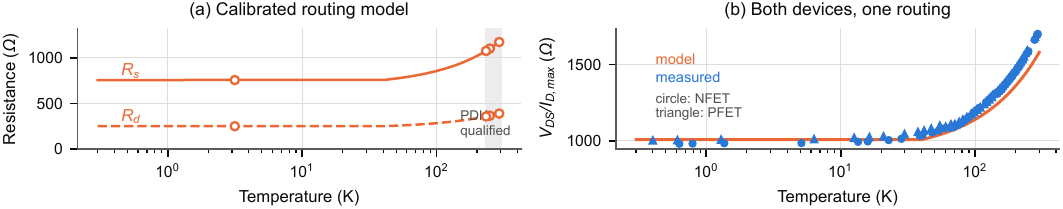}
    \caption{Calibrated source and drain routing
    resistance against temperature. (b) The same model against
    $V_{DS}/I_{D,max}$ measured on both devices measured across temperature.}
    \label{fig:routing}
\end{figure*}

\noindent \textbf{Deembedding using Interconnect Models over Temperature} 
In the test setup, the interconnection to the devices consisted of on-chip routing to wire-bond pads, and external structures (e.g., PCB traces and wiring). Hence, deembedding is required to move the measurement reference point toward the device terminals. %Both the simulation results and the measured data were used to correct the results. 
The on-die measurements did not contain through structures that could be measured in the cryostat with direct access. However, both the n- and p-FinFET devices used identical on-die pad interconnect and the \ac{PDK} enabled extraction down to 233~K. Therefore, the chip structure's parasitics were extracted between ambient and 233~K. For the remainder of the temperature range down to 10~\ac{mK}, interconnect modeling was used that was performed according to the Matthiessen rule \cite{interconnect}, where the resistivity is saturated to its residual value below 40~K. (Fig. \ref{fig:routing}). In the absence of a dedicated \emph{through} structure, the shared parasitic routing was extracted and de-embedded by leveraging the independent measurements of the n- and p-FinFETs, which utilized identical interconnects. An example of the $I_{DS}$ characteristics with respect to voltage at 5.10~K is shown before and after parasitic de-embedding in \cref{fig:output}. 
\section{Cryogenic Extraction and Compact Modeling}
\subsection{Extraction of Intrinsic Parameters by Fixed-Current Method}
Cryogenic cooling disproportionately impacts contact and routing resistance relative to intrinsic channel properties. At cryogenic temperatures, the %M1-M3/C4 
metal stack resistance limits the external drive, dominating the series path (\Cref{fig:output}). Consequently, extraction methodologies relying on peak-$g_m$ become difficult to apply. Because intrinsic $g_m$ increases faster upon cooling than metal resistance $R_S$ decreases, the $g_m\times R_S$ product is inflated at cryogenic temperatures. To mitigate this interconnect limitation, we apply a fixed-current extraction method evaluated in the weak-to-moderate inversion regime ($3.26~\mu\mathrm{A}/\mu\mathrm{m}$). Evaluating parameters at lower currents holds the $I_D\times R_S$ drop small and, critically, identical for both devices, so the comparison is made under matched and quantified resistance loading rather than in the resistance-limited regime. At peak $g_m$ the $g_m\times R_S$ product reaches 0.999 (N) and 0.962 (P). The extraction sits at the ceiling where external $g_m$ reports $1/R_S$ rather than the device. At the fixed current it stays below 0.72 and 0.64. The apparent N/P enhancement spread narrows correspondingly, from 13\% to 9\%.
\subsection{Cryogenic Subthreshold Characterization}
Weak-inversion metrics are largely independent of the series resistance artifact, allowing for reliable subthreshold characterization \cite{Oka2023_iedm}. Both n- and p-FinFET devices avoid carrier freeze-out down to 10~mK, maintaining gate control suitable for \ac{RF-DAC} switches. The intrinsic \ac{SS} floors, taken as the mean voltage change over a 6.5--65~nA/\textmu m decade and averaged below 20~K, are 24.8~mV/dec (N) and 48.4~mV/dec (P). Both floors lie far above the thermal limit $\ln(10)k_BT/q$ and saturate below ${\sim}40$~K, as reported for cryogenic FinFETs~\cite{Chabane2021_esscirc}. The extracted value depends on the current decade it is fitted over: on the intrinsic axis the NFET reads 24.8, 31.4 and 36.1~mV/dec over the 0.1--1, 1--10 and 10--100~\textmu A decades. Series resistance is not the cause, because it is de-embedded first; on the external axis the same decades read 25.5, 38.2 and 104.2~mV/dec. What remains is the weak-to-moderate inversion transition being sampled at successively higher drive. Fin-to-fin $V_{th}$ dispersion across the 320 fins broadens that transition in gate voltage: parallel fins of equal slope sum to the same slope in weak inversion, but they leave it at different gate biases, so the composite curve departs from exponential at a lower current density than a single fin would. An \ac{SS} floor is therefore comparable between devices only at matched current density, and is quoted here with the decade it was fitted over.
\subsection{Decoupled Compact Modeling}
Using the extracted low-current metrics, we developed a continuous, temperature-scalable compact model qualified from 1.36~K to 292~K. To simulate these devices without invalidating the room-temperature PDK qualification, cryogenic physical effects are dynamically injected via external parameter hooks (\texttt{p\_vta} and \texttt{u0mult\_fet}). The threshold shift and mobility correction terms evaluate to zero at 298.15~K, enabling independent tuning of room-temperature levels and cryogenic slopes. This model captures the thermal saturation trends that nominal \ac{PDK} models omit. For the NFET, the model yields RMS errors of 1.7\% for $V_{th}$, 5.2\% for \ac{SS} and 2.3\% for the $R_{on}$ factor; the PFET yields 1.0\%, 10.5\% and 2.2\% respectively. The larger PFET \ac{SS} error reflects the measurement as much as the model: the p-type subthreshold slope resolves at only seven temperatures below 21~K and scatters over 39.5--61.6~mV/dec there, so the measurement uncertainty is comparable to the residual being quoted and this comparison cannot separate the two. 
\begin{table}[!t]
\caption{Macro-model qualification vs. de-embedded measurements}
\label{tab:qualification}
\centering
\footnotesize
\begin{tabular}{llrrr}
\hline
\textbf{Device} & \textbf{Metric} & \textbf{RMS error} & \textbf{Points} & \textbf{Range (K)} \\
\hline
\multirow{4}{*}{NFET} & $|V_{th,CC}|$ & 1.7\,\% & 46 & 1.36--292 \\
& SS & 5.3\,\% & 43 & 1.36--277 \\
& $R_{on}$ factor & 2.3\,\% & 46 & 1.36--292 \\
& $g_m$ @ 50\,\textmu A & 2.1\,\% & 46 & 1.36--292 \\
\hline
\multirow{4}{*}{PFET} & $|V_{th,CC}|$ & 1.0\,\% & 43 & 1.29--279 \\
& SS & 13.2\,\% & 27 & 1.29--244 \\
& $R_{on}$ factor & 2.2\,\% & 43 & 1.29--279 \\
& $g_m$ @ 50\,\textmu A & 9.0\,\% & 43 & 1.29--279 \\
\hline
\multicolumn{5}{l}{\textit{Cryogenic $g_m$ enhancement, $g_m(T_{min})/g_m(T_{max})$:}} \\
\multicolumn{2}{l}{NFET} & \multicolumn{3}{l}{measured 1.89$\times$, model 1.74$\times$} \\
\multicolumn{2}{l}{PFET} & \multicolumn{3}{l}{measured 1.70$\times$, model 1.52$\times$} \\
\hline
\end{tabular}
\end{table}

\section{Conclusion}

\ac{GF} 12LP FinFETs exhibit robust sub-Kelvin gate control suitable for \ac{RF-DAC} applications, with intrinsic \ac{SS} floors of 24.8~mV/dec (N) and 48.4~mV/dec (P). We demonstrated that traditional peak-$g_m$ extraction at deep-cryogenic temperatures is compromised by metal-stack $R_S$, particularly for large multi-fin, multi-finger devices, where $g_m\times R_S$ approaches unity and external $g_m$ saturates at $1/R_S$, necessitating a fixed-current methodology. By injecting these isolated intrinsic metrics via external parameter hooks, we achieved an accurate, temperature-continuous compact model that preserves room-temperature \ac{PDK} baselines.
\onecolumn
% Funding sits here rather than in the title \thanks, as IEDM does; the
% title footnote carries only the equal-contribution note.
\section*{Acknowledgment}
The 12nm finFET IC fabrication was provided by GlobalFoundries as part of their University Partnership Program.

\printbibliography

\begin{figure*}[htbp]
    \centering
    \includegraphics[width=\textwidth]{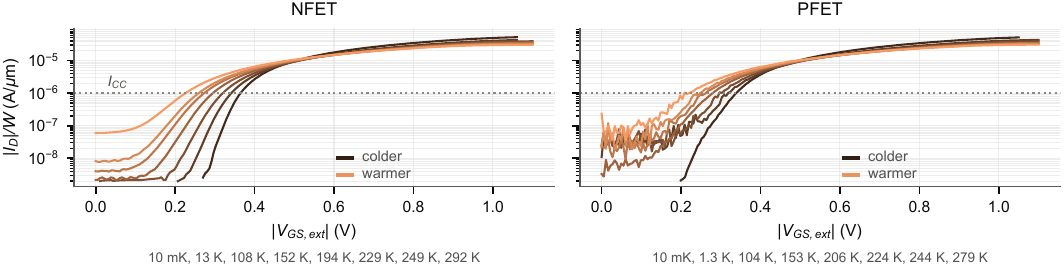}
    \caption{Measured transfer characteristics at $|V_{DS,ext}| = 0.8$~V,
    shaded dark to light with increasing temperature. Drain current is
    normalised to gate width, so the dotted line marks the constant-current
    threshold criterion at $I_{CC} = 1$~\textmu A/\textmu m.}
    \label{fig:transfer}
\end{figure*}

\vspace{-100mm}
\begin{figure*}[htbp]
    \centering
    \includegraphics[width=\textwidth]{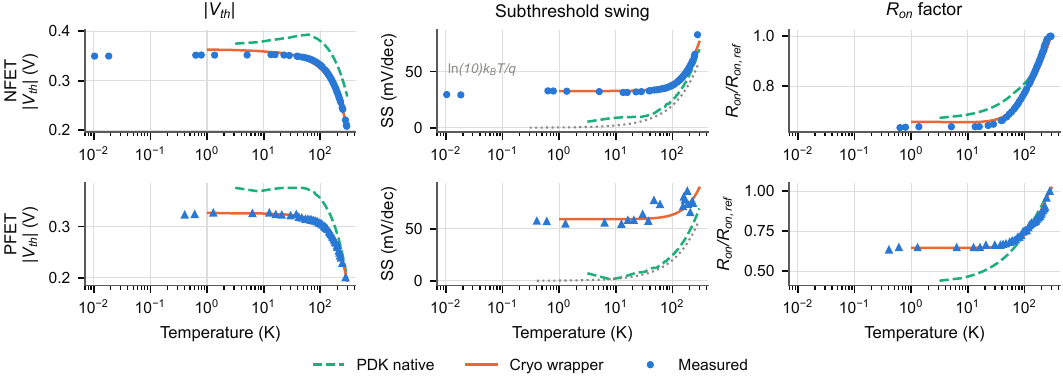}
    \caption{De-embedded parameters against temperature: $|V_{th}|$, SS and $R_{ON}$.}
    \label{fig:params}
\end{figure*}

\begin{figure*}[htbp]
    \centering
    \includegraphics[width=\textwidth]{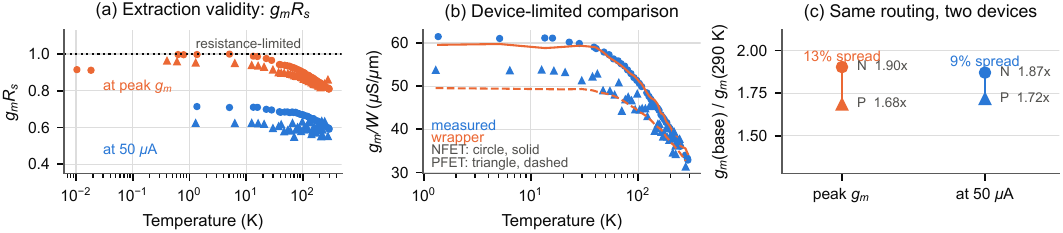}
    \caption{(a)$gm\times R_S$ saturates at cryo temperatures because $g_m$ increases faster than $R_S$ decreases. (b)~$g_m/W$ (extracted at 50~\textmu A) compared to the proposed wrapper model for n- and p=FinFETs. (c)~Ratio of $g_m$ at peak and minimum temperature before and after de-embedding for n- and p-FinFET.}
    \label{fig:gm}
\end{figure*}

\end{document}